# Адаптація дизайну сайту для людей із порушенням кольоросприйняття

Анна Михайлівна Горло[*], Ірина Сергіївна Мінтій[#]
Кафедра інформатики та прикладної математики,
Криворізький державний педагогічний університет,
пр. Гагаріна, 54, м. Кривий Ріг, 50086, Україна
stifa1996@gmail.com[*], irina.mintiy@gmail.com[#]

**Анотація**. *Метою дослідження* є опис проблеми розробки веб-дизайну для людей з порушеннями кольоросприйняття. *Задачами дослідження* є ознайомлення з існуючими алгоритмами імітації порушень кольоросприйняття та пошук найбільш відповідних колірних моделей для реалізації фільтрування спірних кольорів. *Об'єктом дослідження* є конвертація колірних моделей та алгоритм фільтрування. *Предметом дослідження* є методи визначення спірних кольорів. У статті досліджено проблеми людей із різними видами порушень кольоросприйняття, розглянуто основні положення теорії трикомпонентного зору, обгрунтовано необхідність переходу з одних колірних моделей до інших, наведено формули конвертації з RGB-моделі до HSL-моделі, систематизовано алгоритми імітації та фільтрування кольорів у різних видах дихроматизму: протанопії, дейтеранопії та тританопії. *Результати дослідження* планується використати при розробці системи адаптації дизайну сайту для людей із порушеннями кольоросприйняття.

**Ключові слова**: дальтонізм; колірна модель; веб-дизайн.

**A. M. Horlo[*], I. S. Mintii[#]. Adapting website design for people with color-blindness**

**Abstract.** *The aim of the study* is the description of problem of developing web design for people with color blindness. *The objectives of the study* are familiarising with the exiting algorithms of simulation color blindness and searching the most appropriate color models to realize a filter of disputed colors. *The object of the study* is the convertation of color models and algorithms of filtration. *The subject of the study* are methods of recognition disputed colors. In the study were investigated the problems of color blind people, examined the basic concepts of trichromatic color vision theory, substantiated the necessity of changing different types of color models, given formulas convertation from RGB-color model to HSL-color model, systematized the algorithms of imitation and filtration of colors for different types of dichromacy: protanopia, deuteranopia and tritanopia. *The results of the* *study* are planned using in development of adapting website design for people with color blindness.

**Affiliation**: Department of Computer Science and Applied Mathematics, Kryvyi Rih State Pedagogical University, 54, Gagarin Ave., Kryvyi Rih, 50086, Ukraine.

E-mail: stifa1996@gmail.com[*], irina.mintiy@gmail.com[#].

Згідно [4], «інклюзивне навчання має покращити навчальне середовище, забезпечити потреби всіх учнів з повагою до їхніх здібностей та можливостей». Враховуючи, що у світі близько 5-8 % чоловіків і 0,5 % жінок мають колірну сліпоту (дальтонізм), дуже важливо при проектуванні дизайну сайту врахувати проблеми людей з аномаліями кольоросприйняття.

Згідно трикомпонентної теорії кольорового зору (теорія Юнга-Гемгольца), всі порушення кольоросприйняття пов'язані з відсутністю пігменту тих чи інших колбочок [1].

Відомі три типи дефектів кольорового зору: аномальний трихроматизм, дихроматизм і монохроматизм.

Особам з аномальним трихроматизмом для сприйняття кольору потрібні інші кількості основних кольорів, ніж людині без порушень. Як правило, аномальний трихроматизм проявляється у формі протаномалії і дейтераномалії [3].

У протаномада – людини, що страждає протаномалією, – дефіцит пігменту L-колбочок, у наслідок чого він недостатньо чутливий до червоних тонів, властивих довгохвильовому світлу.

У дейтераномалії знижена чутливість до зелених тонів, характерних для середньохвильового світла, що є результатом нестачі пігменту M-колбочок.

Протаномаду потрібно більше червоного, ніж людині з нормальним кольоросприйняттям, а дейтераномаду – більше зеленого.

Людям, які страждають дихроматизмом, для відтворення всіх відтінків кольорів потрібні всього два кольори, а не три, як потрібно людям з нормальним кольоровим зором. Різновидами дихроматизму є дейтеранопія, протанопія та тританопія, що виявляються в різкому зниженні чутливості до зеленого, червоного і синього кольорів відповідно.

Протанопія – порушення сприйняття червоного кольору. Люди з таким порушенням сприймають червоний колір темнішим, він поєднується з темно-зеленим, темно-коричневим, а зелений – зі світло-сірим, жовтим, світло-коричневим [2].

Дейтеранопія – порушення сприйняття зеленого кольору. У людей з таким порушенням зелений колір поєднується зі світло-помаранчевим, рожевим, а червоний – з зеленим та світло-коричневим [2].

Рідкісна третя форма дихроматизму, так звана тританопія, є результатом відсутності пігменту S-колбочок. Тританопи погано розрізняють синій і жовтий кольори, вони бачать лише червоний і зелений і плутають жовті, сірі і сині відтінки [2].

Монохроматизм – повна відсутність можливості сприймати будь який інший колір, крім білого, чорного та відтінків сірого [2].

Для того, аби розробляти версії дизайну сайту для людей із дальтонізмом, необхідно спочатку змоделювати різні можливі види дальтонізму. Для цього існує спеціальний алгоритм, розроблений для колірної моделі LMS.

У сітківці знаходяться колбочки трьох видів, кожна із яких має чутливість до світла певної довжини. Дані колбочки мають відповідну назву S, M, L залежно від довжини хвилі світла. Для трикомпонентного зору існує колірна модель, що найкращим чином відтворює принцип сприймання кольору людським оком. Ця колірна модель називається LMS. Але для подальших розрахунків і відображення результатів необхідно використовувати такі кольорові схеми, які наявні у веб-просторі. Для цієї цілі може підійти RGB модель, оскільки вона представляє собою відображення трьох основних кольорів, які сприймає наше око (red, green, blue). Тобто, спочатку необхідно виконати конвертацію з RGB-моделі до LMS-моделі за допомогою наступної матриці [1]:

$$\begin{pmatrix} L \\ M \\ S \end{pmatrix} = T_{RGB-LMS} * \begin{pmatrix} R \\ G \\ B \end{pmatrix}.$$

Наступним кроком необхідно виконати перетворення нормальних значень LMS-моделі до значень з різними видами дальтонізму.

Для протанопії: $\begin{pmatrix} L_p \\ M_p \\ S_p \end{pmatrix} = \begin{pmatrix} 0 & 2.0234 & -2.5258 \\ 0 & 1 & 0 \\ 0 & 0 & 1 \end{pmatrix} \begin{pmatrix} L \\ M \\ S \end{pmatrix}$

Приклад імітації протанопії наведено на рис. 1 (б).

Для дейтеранопії: $\begin{pmatrix} L_d \\ M_d \\ S_d \end{pmatrix} = \begin{pmatrix} 1 & 0 & 0 \\ 0.4942 & 0 & 1.2483 \\ 0 & 1 & 1 \end{pmatrix} \begin{pmatrix} L \\ M \\ S \end{pmatrix}.$

Приклад імітації дейтеранопії наведено на рис. 1 (в).

На останньому етапі необхідно здійснити конвертацію даних з LMS-моделі до RGB-моделі.

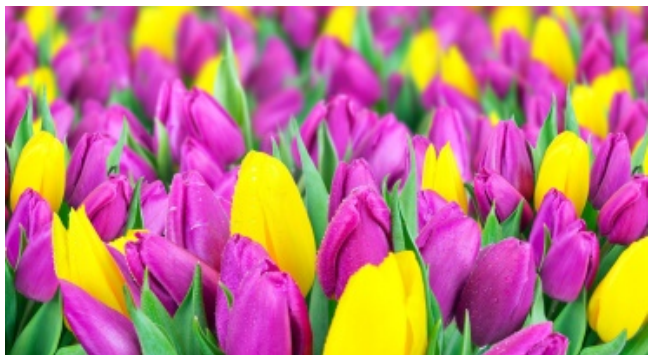

*а)*

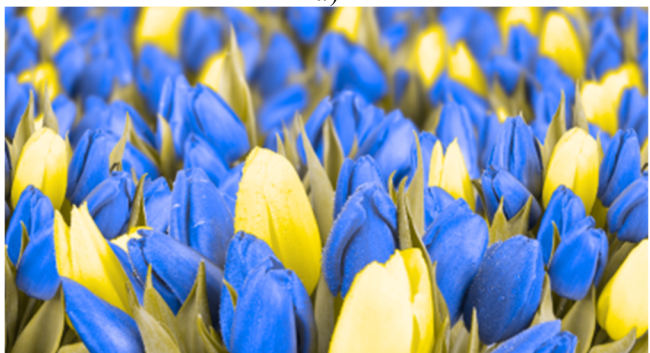

*б)*

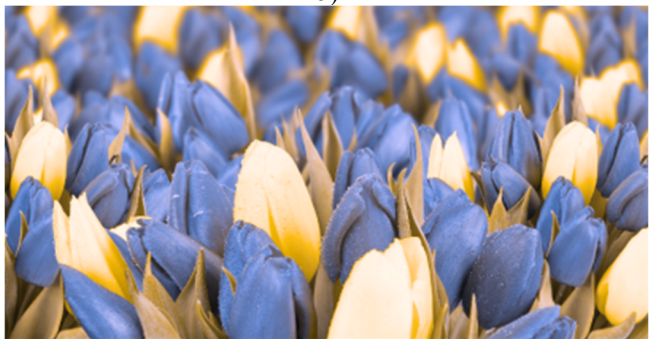

*в)*

Рис. 1. а) зображення для людини без дальтонізму; б) зображення для людини із протанопією; в) зображення для людини із дейтеранопією

Після моделювання необхідних кольорів можна зрозуміти, чи важким для сприйняття буде дизайн майбутнього сайту для людей із порушенням кольоросприйняття.

Принцип фільтрації помилок такий: якщо html-об'єкти знаходяться близько один біля одного і мають фонові кольори або колір тексту такі, що будуть конфліктувати у людини з дальтонізмом, необхідно такі кольори замінити на інші кольори із допустимого спектру.

Для того, аби замінити колір, необхідно звернутися до HSL-моделі.

HSL-модель має більш логічне уявлення кольору, ніж RGB. Дана колірна модель представлена трьома характеристиками: насиченістю, кольоровим тоном та світлотою (яскравістю). Таку колірну модель дуже зручно використовувати для фільтрації спектру, оскільки вона представлена у вигляді кола (рис. 3).

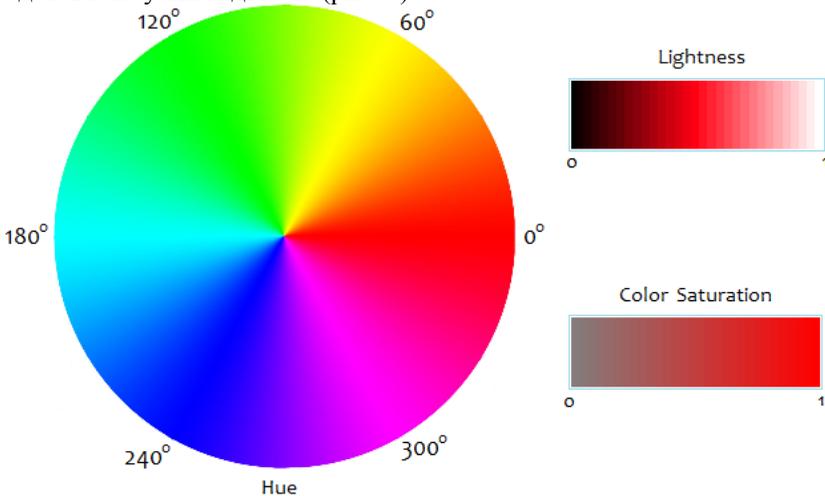

Рис. 3. Колірна модель HSL

Через це дуже зручно знаходити протилежний колір спектру. Для того, аби перейти з RGB-моделі до HSL-моделі, необхідно виконати конвертацію, використовуючи наступні формули:

$$H = \begin{cases} 60° * \frac{G-B}{MAX-MIN} + 0°, & if\ MAX = R\ and\ G \geq B \\ 60° * \frac{G-B}{MAX-MIN} + 360°, & if\ MAX = R\ and\ G < B \\ 60° * \frac{G-B}{MAX-MIN} + 120°, & if\ MAX = G \\ 60° * \frac{G-B}{MAX-MIN} + 240°, & if\ MAX = B \end{cases}$$

$$S = \frac{MAX-MIN}{1-|1-(MAX+MIN)|},$$

$$L = \frac{1}{2}(MAX + MIN),$$

де: MAX – максимум з трьох значень (R, G, B); MIN – мінімум з трьох значень (R, G, B); H – тон [0; 360]; S – насиченість [0; 1]; L – яскравість (світлота) [0; 1].

Після імітації порушень кольоросприйняття та фільтрації кольорів можна переходити до проектування та розробки системи адаптації дизайну сайту для людей з порушенням кольоросприйняття, що є напрямом подальших досліджень.